\documentclass[prd,twocolumn,aps,superscriptaddress,showpacs]{revtex4}
\usepackage{amssymb}
\usepackage{amsmath,bm}
\usepackage{graphicx}
\usepackage[normalem]{ulem}
\usepackage[dvips]{color}
\usepackage{subfigure}
\newcommand{\be}{\begin{equation}}

\newcommand{\ee}{\end{equation}}
\newcommand{\bea}{\begin{eqnarray}}
\newcommand{\eea}{\end{eqnarray}}

\renewcommand\sout{\bgroup \color{red} \ULdepth=-.5ex \ULset}

\begin{document}

\title{Warm asymmetric quark matter and proto-quark stars within the confined-isospin-density-dependent mass model }
\author{Peng-Cheng Chu}
\email{kyois@126.com}
\affiliation{Qingdao Technological University, Qingdao 266000, China}

\author{Lie-Wen Chen}
\email{lwchen@sjtu.edu.cn}
\affiliation{Department of Physics and Astronomy and Shanghai Key Laboratory for Particle
Physics and Cosmology, Shanghai Jiao Tong University, Shanghai 200240, China}
\affiliation{Center of Theoretical Nuclear Physics, National Laboratory of Heavy-Ion
Accelerator, Lanzhou, 730000, China}


\begin{abstract}
We extend the confined-isospin-density-dependent mass (CIDDM) model to include temperature dependence of the equivalent mass for quarks. Within the CIDDM model, we study the equation of state (EOS) for $\beta$-equilibrium quark matter, quark symmetry energy, quark symmetry free energy, and the properties of quark stars at finite temperatures. We find that including the temperature dependence of the equivalent mass can significantly influence the properties of the strange quark matter (SQM) as well as the quark symmetry energy, the quark symmetry free energy, and the maximum mass of quark stars at finite temperatures. The mass-radius relations for different stages of the proto-quark stars (PQSs) along the star evolution are analyzed. Our results indicate that the heating (cooling) process for PQSs will increase (decrease) the maximum mass within the CIDDM model by including temperature dependence of the equivalent mass for quarks.
\end{abstract}

\maketitle

\section{Introduction}

The investigation of the properties of strong interaction matter, as one of the fundamental issues in contemporary nuclear physics, astrophysics, and cosmology, plays a central role in understanding the nuclear structures and reactions, the critical issues in astrophysics, and the matter state at early universe. In terrestrial laboratories, the experiments of high energy heavy ion collisions(HICs) provide a unique tool to explore the properties of strong interaction matter, and the hot quark-gluon plasma is expected to be created in HICs at the Relativistic Heavy Ion Collider(RHIC) at BNL and the Large Hadron Collider(LHC) at CERN. The hot and dense quark matter might be created in HICs from the Facility for Antiproton and Ion Research(FAIR) at GSI and the Nuclotron-based Ion Collider Facility(NICA) at JINR, while the cold and dense quark matter may exist in the inner core of compact stars. Compact stars, in nature, provide a unique astrophysical testing ground to explore the properties of the strong interaction matter at high baryon density and low temperature~\citep{Gle00,Web99}. Neutron stars~(NSs) are a class of the densest compact stars in the Universe and have been shown to provide the natural testing grounds of our knowledge about the EOS of neutron-rich nuclear matter~\citep{Lattimer04,Steiner05}. In the interior of NSs, the baryon density can reach or even be larger than about six times the normal nuclear matter density $n_0=0.16~\text{fm}^{-3}$, so there might exist hyperons, meson condensations, and even quark matter. Theoretically, NSs may be converted to (strange) quark stars (QSs)~\citep{Bom04,Sta07,Her11}, which is composed of absolutely stable deconfined $u$, $d$, and $s$ quark matter~(with some leptons), i.e., strange quark matter (SQM).  Although most observations related to compact stars can be explained by the conventional NS models, the QS hypothesis cannot be conclusively ruled out.
The possible existence of QSs is one of the most intriguing aspects of modern astrophysics and has important implications for studying the strong interaction matter physics, especially the properties of SQM which essentially determine the structure of QSs~\citep{Iva69,Ito70,Bodmer71,Witten84,Far84,Alc86,Web05}.
{The detection of the gravitational waves which is emitted by the oscillation for a nonspherical supernova explosion may be a useful way to obtain the information for quark stars~\citep{Sotani03,Sotani04}, and one may discriminate QSs from NSs in quadrupole f-mode oscillation, which is one of the most important aspect for gravitational radiation~\citep{Kojima02}.}

The type II supernova explosion is triggered when massive stars exhaust the fuel supply, which will cause the core to be crushed by gravity. The result of this gravitational collapse is to form a neutron star or a black hole, which depends on the initial condition of the collapse. {Since SQM may be the true ground state of matter~\citep{Iva69,Ito70,Bodmer71,Witten84}, neutron stars can be converted to QSs as soon as the core of the star converts to the quark phase, and there might be only QSs instead of neutron stars~\citep{Alc86}. The formation of proto neutron stars~(PNS) can be relatively well understood in the work\citep{Prakash1997}, while the transition from PNS to proto-quark stars~(PQS) during the type II supernova explosion is known little owing to the lack of detailed understanding of the complex burning process of hadron matter into SQM. Several works are done to investigate the existence of possible stable SQM and the properties of PQS~\citep{Alc86, Gupta03, Su05,Menezes13,Menezes Influnece} without considering the scenario how to produce PQSs. From the results in the work~\cite{Drago14, Drago16, Bauswein16}, PQSs could be produced from the merging of neutron stars, and for PQS there might be a dynamically unimportant crust of nucleonic matter with densities below the neutron drip density. }

{We discuss the temperature effects on the properties of SQM and PQSs and neglect the possible crust structure for PQSs in the present work.} In the newly-born proto-quark stars, neutrinos are produced by electron capture in SQM and trapped due to their short mean free paths from leaving the star on dynamical timescales. At the very beginning stage of the birth of a PQS, the lepton number per baryon with these trapped neutrinos is approximately 0.4, which depends on the efficiency of electron capture reactions during the gravitational collapse of the progenitor star, and the entropy per baryon is about one. During the following 10-20 seconds, the star matter will be heated by the diffusing neutrinos, and the entropy per baryon will increase to two, while the neutrino fraction decreases almost down to zero. Following this heating stage, PQSs begins cooling down, and finally forms into the cold QSs. Along the evolution line from a PQS to a cold QS, people usually take the snapshots to study how the star evolves as $\text{(I)}~S/n_B=1, Y_l=0.4;~\text{(II)}~
S/n_B=2, Y_{\nu_l}=0;~\text{(III)}~
S/n_B=0, Y_{\nu_l}=0$~\cite{SteinerPLB00,Shao35,Menezes Influnece, Menezes PNS, Shao11}, where $S$ is the entropy per baryon for star matter, $n_B$ is the baryon density, and $Y_l$ and $Y_{\nu_l}$ stand for the lepton fraction at extreme isospin and neutrino fraction respectively.

In QSs, the star matter has large $u$-$d$ quark asymmetry~(isospin asymmetry), which indicates that the iso vector properties of SQM may play an important role, and the quark matter formed in high energy HICs at RHIC/LHC generally has unequal $u$ and $d$ ($\bar{u}$ and $\bar{d}$) quark numbers, which is also isospin asymmetric. In the previous studies of PQSs and proto-hybrid stars~\cite{Steiner01,SteinerPRL01,Ni06,Burgio08,YA09,Chen2012} , the large isospin asymmetry can still be found in star matter, which can also be related to the isovector properties of quark matter at finite temperature and finite baryon density. Therefore, it is of great interest and critical importance to explore the isovector properties of the warm asymmetric quark matter, which is very useful for understanding the properies of PQSs, the QCD phase diagram at extreme isospin, and the isospin effects of partonic dynamics in high energy HICs.

In the present work, we extend the confined-isospin-density-dependent mass~(CIDDM) model to include temperature dependence of the equivalent mass for quarks to investigate the quark matter symmetry energy/ symmetry free energy and the properties of SQM and PQSs. We find that the temperature dependence of the equivalent quark mass introduced in the CIDDM model can significantly influence the EOS of SQM, the values of symmetry energy/ symmetry free energy, and the maximum mass for QSs at finite temperature and PQSs in the three snapshots along the star evolution.

\section{MODELS AND METHODS}
\subsection{The confined isospin- and density-dependent mass model}
\label{Sec-CIDDM}
The confined isospin- and density-dependent mass ( i.e., the  CIDDM) model~\cite{Chu2014, Chu2014b, Chu2014c}
is an extended version of the CDDM model~\cite{Fow81,Cha89,Cha91,Ben95,Pen99,Peng00,Peng08,Li11}
for asymmetric quark matter by including the isospin dependence of the equivalent
mass for quarks. In the CIDDM model, density- and
isospin-dependent quark masses are used to describe the quark confinement. With baryon number density $n_B$ and isospin
asymmetry $\delta $, the equivalent mass is expressed as
\begin{eqnarray}
m_q &=& m_{q_0} + m_{I} + m_{iso} \notag \\
&=& m_{q_0} + \frac{D}{{n_B}^z} - \tau_q \delta {D_I}n_B^{\alpha}e^{-\beta n_B},
\label{mqiso}
\end{eqnarray}
where $m_{q0}$ is the quark current mass, $m_I = \frac{D}{{n_B}^z} $ represents
the flavor-independent quark interactions, while
$m_{iso} = - \tau_q \delta {D_I}n_B^{\alpha}e^{-\beta n_B}$  reflects the
isospin dependent part. For $m_I = \frac{D}{{n_B}^z} $,
the constant $z$ is the equivalent mass scaling parameter and the constant $D$ is a
parameter determined by stability arguments of SQM. For
$m_{iso} = - \tau_q \delta {D_I}n_B^{\alpha}e^{-\beta n_B}$, the constants $D_I$,
$\alpha $ and $\beta $ are parameters determining the isospin-density dependence of the
effective interactions in quark matter, $\tau_q $ is the isospin quantum number of quarks, and we set $\tau_q = 1$ for $q=u$ ($u$ quarks), $\tau_q = -1$ for $q=d$
($d$ quarks), and $\tau_q = 0$ for $q=s$ ($s$ quarks). From the works~\cite{DiT06,Pag10,DiT10,Sha12,Chu2014},
the isospin asymmetry is defined as
\begin{equation}
\delta = 3\frac{n_d-n_u}{n_d+n_u},
\label{delta}
\end{equation}
which equals to $-n_3/n_B$ with the isospin density $n_3 = n_u-n_d$ and
$n_B = (n_u+n_d)/3$ for two-flavor $u$-$d$ quark matter. We can find that one has
$\delta = 1$ ($-1$) for quark matter converted by pure neutron (proton) matter
according to the nucleon constituent quark structure, which is consistent with the
conventional definition for nuclear matter, i.e.,
$\frac{n_n -n_p}{n_n +n_p}=-n_3/n_B$.

In Eq. (\ref{mqiso}), if the scaling parameter $z > 0$ and $\alpha \ge 0$, one can find that the quark confinement condition
$\lim_{n_B\to0}m_q=\infty$ will be guaranteed.
In addition, if $\beta >0$, then $\lim_{n_B \to \infty} m_{iso} = 0$
and thus the asymptotic freedom $\lim_{n_B\to \infty} m_q=m_{q0}$ is satisfied.
For two-flavor $u$-$d$ quark matter, the chiral symmetry is restored at high
density due to $\lim_{n_B\to\infty}m_q = 0$ if the current masses of $u$ and $d$
quarks are neglected.  Therefore, the phenomenological
parametrization form of the isospin dependent equivalent quark mass in
Eq.~(\ref{mqiso}) is very general and respects the basic features of QCD.
For more details about the CIDDM model, the readers are referred to Ref.~\cite{Chu2014}.
  As demonstrated in Ref.~\cite{Wen05} and Ref.~\cite{Zhang03}, temperature dependence of equivalent mass in the CDDM model can be obtained by considering the linear confinement and string tension $\sigma(T)$. We introduce the temperature dependence of equivalent mass for quarks in the CIDDM model by using the similar way as in Ref.~\cite{Wen05} and Ref.~\cite{Zhang03}, and the equivalent mass is modified as

  \begin{eqnarray}
 m_q=m_{q_0}+\left(\frac{D}{{n_B}^{z}}-\tau_q\delta{D_I} n_B^{\alpha}e^{-\beta n_B}\right)\sigma(T),
\end{eqnarray}
with
\begin{eqnarray}
\sigma(T)=1-\frac{8T}{\lambda T_c}\exp\left(-\lambda \frac{T_c}{T}\right),
\end{eqnarray}
  where $q=u,d,s$, $\sigma(T)$ is the temperature dependent string tension \cite{Ukawa93}, $T_c=170\text{MeV}$ is the critical temperature {(which is calculated from LQCD \cite{fodor02}, and this value is used in Ref. \cite{Wen05} and Ref. \cite{Zhang03} for temperature dependence)}, and $\lambda= 1.605812$ is determined as the solution of the equation $1-\frac{8T}{\lambda T_c}\exp(-\lambda \frac{T_c}{T})=0 $ when $T=T_c$.
  One can also find that this isospin-density-temperature dependent equivalent mass will turn back to be the same form as Eq.~(\ref{mqiso}) once zero temperature case is considered. For the finite temperature case, $m_q$ will decrease as temperature increases. When the temperature hits the critical value $T_c$, this isospin-density-temperature dependent equivalent mass will reach the quark current mass $m_{q0}$, which models the chiral symmetry restoration.

\subsection{The quark matter symmetry energy and symmetry free energy}
The EOS of quark matter consisting of $u, d,$ and $s$ quarks, which is defined by its binding energy per baryon number, can be expanded in isospin asymmetry $\delta$ as
\begin{equation}
E(n_B ,\delta, n_s)=E_{0}(n_B, n_s)+E_{\mathrm{sym}}(n_B, n_s)\delta ^{2}+\mathcal{O}(\delta ^{4}),
\label{esym}
\end{equation}
where $E_{0}(n_B, n_s)=E(n_B ,\delta =0, n_s)$ is the binding energy per baryon number in three-flavor $u$-$d$-$s$ quark matter with an equal fraction of $u$ and $d$ quarks. One can find that Eq.~(\ref{esym}) is similar to the case of nuclear matter\cite{Li08} and the quark matter symmetry energy is expressed as
 \begin{eqnarray}
E_{\mathrm{sym}}(n_B, n_s) &=&\left. \frac{1}{2!}\frac{\partial ^{2}E(n_B
,\delta, n_s)}{\partial \delta ^{2}}\right\vert _{\delta =0}.
\label{binding}
\end{eqnarray}
In order to investigate the thermodynamical properties for asymmetric quark matter, one should also consider the symmetry free energy in $u$-$d$-$s$ quark matter at finite temperature. Since using the similar expression as symmetry energy is a feasible way of defining symmetry free energy, the quark matter symmetry free energy $F_{sym}$ can be expressed as
 \begin{eqnarray}
{F}_{\mathrm{sym}}  &=&\left. \frac{1}{2!}\frac{\partial ^{2}F}{\partial \delta ^{2}}\right\vert _{\delta =0},
\label{symfree}
\end{eqnarray}
where $F$ is the free energy per baryon for quark matter. One can find that symmetry free energy and symmetry energy have the same value at zero temperature, because the entropy density is zero and the values of free energy and energy will be identical, while there should be a temperature dependent difference between symmetry free energy and symmetry energy once finite temperature condition is considered.
\subsection{Properties of strange quark matter at finite temperature}

The EOS is the most important aspect for physicists to obtain
the properties of quark matter. For SQM, we assume it is composed of $u$, $d$, and $s$ quarks and $e$, $\mu$, $\nu_{e}$ and $\nu_{\mu}$ leptons with electric charge neutrality in beta-equilibrium. The weak beta-equilibrium condition can be expressed as
 \begin{align}
\mu_d=\mu_s=\mu_u+\mu_e-\mu_{\nu_e},\\ \mu_{\mu}=\mu_e ~~~\text{and}~~~\mu_{\nu_{\mu}}=\mu_{\nu_e}.
\label{beta}
\end{align}
Furthermore, the electric charge neutrality condition can be written as
\begin{eqnarray}
\frac{2}{3}n_u=\frac{1}{3}n_d+\frac{1}{3}n_s+n_e+n_{\mu}.
\end{eqnarray}

{The quasiparticle contribution to the total thermodynamic potential density for SQM can be written as
\begin{eqnarray}
\Omega=\sum_i\Omega_i,
\end{eqnarray}
where $i$ in the sum is for all flavors of quarks ($u$, $d$, and $s$) and leptons ($e$, $\mu$, $\nu_{e}$, and $\nu_{\mu}$). Then the contribution of the particle with flavor i to the thermodynamic potential density is
\begin{eqnarray}
 \Omega_i = &-&\frac{g_iT}{2\pi^2}\int_0^\infty\{\ln[1+e^{-(\epsilon_i-\mu_i^*)/T}]\notag\\
       ~&+&\ln[1+e^{-(\epsilon_i+\mu_i^*)/T}]\}p^2\text{d}p,
\label{Omega}
\end{eqnarray}
  where $\epsilon_i = \sqrt{m_i^2+p^2}$ is the corresponding energy-momentum dispersion, and $\mu_i^*$  is the effective chemical potential (For leptons, the form of the effective chemical potential is identical to the form of their chemical potential). The value of the degeneracy factor $g_i$ is $6$ for quarks, while $g_i=2$ for leptons. For the particle with flavor i, the number density can be obtained as
\begin{eqnarray}
 n_i = \frac{g_i}{2\pi^2}\int_0^\infty[\frac{1}{1+e^{(\epsilon_i-\mu_i^*)/T}} -\frac{1}{1+e^{(\epsilon_i+\mu_i^*)/T}}]p^2\text{d}p.\notag\\
\label{Omega}
\end{eqnarray}
The free-energy density $\mathcal{F}$ and the pressure $P$ are, respectively,
 \begin{eqnarray}
\mathcal{F}=\sum_i\mathcal{F}_i=\sum_i(\Omega_i+\mu_i^* n_i),\label{Free}\\
P=-\sum_i \Omega_i + \sum_{i,j} \frac{\partial \Omega_j}{\partial m_j}\frac{\partial m_j}{\partial n_i}n_i\label{Pressure},
\label{Omega}
\end{eqnarray}
where the second term in Eq.$~(\ref{Pressure})$ is produced by the density dependence of the equivalent quark mass, and this additional part is important for guaranteeing the thermodynamic self-consistency of the model.}

Using $\mu_i={\text{d}{\mathcal{F}}}/{\text{d}{n_i}}$, the effective chemical potential for $u$ quark at finite temperature can be expressed analytically as
\begin{align}
\mu_u^*=&\mu_u -\Big\{\frac{1}{3}\sum_{j=u,d,s}\frac{\partial \Omega_j}{\partial m_j}\notag\\
&\times\left[-\frac{zD}{n_B^{(1+z)}}- \tau_j D_I\delta(\alpha n_B^{\alpha-1}-\beta n_B^\alpha)e^{-\beta n_B}\right] \notag\\
&+D_In_B^\alpha e^{-\beta n_B}\left(\frac{\partial \Omega_u}{\partial m_u}-\frac{\partial \Omega_d}{\partial m_d}\right) \notag\\
&\times\frac{6n_d}{(n_u+n_d)^2} \Big\}\times\left[1-\frac{8T}{\lambda T_c}exp(-\lambda \frac{T_c}{T})\right].
\label{u}
\end{align}%
For $d$ and $s$ quarks, we have, respectively,
\begin{align}
\mu_d^*=&\mu_d -\Big\{\frac{1}{3}\sum_{j=u,d,s}\frac{\partial \Omega_j}{\partial m_j}\notag\\
&\times\left[-\frac{zD}{n_B^{(1+z)}}- \tau_j D_I\delta(\alpha n_B^{\alpha-1}-\beta n_B^\alpha)e^{-\beta n_B}\right] \notag\\
&+D_In_B^\alpha e^{-\beta n_B}\left(\frac{\partial \Omega_d}{\partial m_d}-\frac{\partial \Omega_u}{\partial m_u}\right) \notag\\
&\times\frac{6n_u}{(n_u+n_d)^2}\Big\}\times\left[1-\frac{8T}{\lambda T_c}exp(-\lambda \frac{T_c}{T})\right],
\label{d}
\end{align}%
and
\begin{align}
\mu_s^*=&\mu_s -\frac{1}{3}\sum_{j=u,d,s}\frac{\partial \Omega_j}{\partial m_j}&\notag\\
&\times\left[-\frac{zD}{n_B^{(1+z)}}- \tau_j D_I\delta(\alpha n_B^{\alpha-1}-\beta n_B^\alpha)e^{-\beta n_B}\right]&\notag\\
&\times\left[1-\frac{8T}{\lambda T_c}exp(-\lambda \frac{T_c}{T})\right].&
\label{s}
\end{align}%

The total energy density is $\mathcal{E}_{}=\sum_i \mathcal{E}_i$ with
  \begin{eqnarray}
 \mathcal{E}_i =   &-&  \frac{g_i}{2\pi^2}\int_0^\infty\left[\frac{\epsilon_{i,p}}{1+e^{(\epsilon_i-\mu_i^*)/T}}
  +  \frac{\epsilon_{i,p}}{1+e^{(\epsilon_i+\mu_i^*)/T}}\right]p^2\text{d}p\notag\\ &-& T\frac{\partial \Omega_i}{\partial m_i}\frac{\partial m_i}{\partial T}\label{energydensity}.
\end{eqnarray}
Then the entropy density and the partial derivatives relevant to $\Omega_i$ in the preceding equations can be calculated as

 \begin{eqnarray}
S=\sum\limits_i S_i =\sum\limits_i(-\frac{\partial\Omega_i}{\partial T}-\frac{\partial\Omega_i}{\partial m_i}\frac{\partial m_i}{\partial T}),\label{entropy}
\end{eqnarray}
\begin{eqnarray}
\frac{\partial \Omega_i}{\partial m_i}= \frac{g_i m_i }{2\pi^2}\int_0^\infty\Big[\frac{1}{1+e^{(\epsilon_i-\mu_i^*)/T}} +\frac{1}{1+e^{(\epsilon_i+\mu_i^*)/T}}\Big]\frac{ p^2}{\epsilon_i}\text{d}p,\notag\\
\end{eqnarray}
and
\begin{eqnarray}
 \frac{\partial{\Omega_i}}{\partial T} = &-&\frac{g_i }{2\pi^2}\int_0^\infty\Big\{\ln{\Big[1+e^{-(\epsilon_i-\mu_i^*)/T}\Big]} +\frac{(\epsilon_i-\mu_i^*)/T}{1+e^{(\epsilon_i-\mu_i^*)/T}}\notag\\&+&\ln{\Big[1+e^{-(\epsilon_i+\mu_i^*)/T}\Big]} +\frac{(\epsilon_i+\mu_i^*)/T}{1+e^{(\epsilon_i+\mu_i^*)/T}}\Big\}p^2\text{d}p.\notag\\
\end{eqnarray}

One can obtain $\mathcal{F}_i=\mathcal{E}_i-TS_i$ by solving Eq.~(\ref{Free}), (\ref{energydensity}) and (\ref{entropy}), which matches the definition of the free energy density and reflects that the values of free energy density and energy density are equal once zero temperature condition is considered.

\subsection{Properties of proto-quark stars}
Using the EOS's of SQM, one can obtain the Mass-radius relation of static QSs by soving the Tolman-Oppenheimer-Volkov (TOV) equation \cite{Oppenheimer39}:
\begin{align}
\frac{dM}{dr}=4\pi r^2 \epsilon(r),
\end{align}
\begin{eqnarray}
\frac{dp}{dr}&=&-\frac{G\epsilon(r)M(r)}{r^2}[1+\frac{p(r)}{\epsilon(r)}]\nonumber \\
& & [1+\frac{4\pi p(r)r^3}{M(r)}][1-\frac{2GM(r)}{r}]^{-1},
\end{eqnarray}
where $M(r)$ is the total mass inside the sphere of radius $r$, $\epsilon(r)$
is the corresponding energy density,
$p(r)$ is the corresponding pressure, and $G$ is Newton's gravitational constant.

For PQS, which cannot be conclusively ruled out to explain the observations of compact stars, the previous studies usually use the similar method to describe the evolution of PQS as that of PNS. At the beginning of the birth of the PQS, the entropy per baryon is about one and the number of leptons per baryon with trapped neutrinos is about $0.4$$(Y_l=Y_e+Y_{\mu}+Y_{\nu_l}=Y_e+Y_{\mu}+Y_{\nu_e}+Y_{\nu_{\mu}}=0.4)$, which is set as the first snapshot of PQS evolution \cite{Steiner01,Shao35}. In the following $10$-$20$ seconds, neutrinos escape from the star and the diffusing neutrinos can heat the star matter \cite{Prakash1997}, which increase the corresponding entropy density. In this stage, the neutrino fraction is almost zero, and one can describe this stage as the second snapshot of PQS evolution. After the heating stage, the star begins cooling by radiating neutrino pairs, then finally a cold quark star forms \cite{Steiner01,Shao35}. In this paper, we describe the time evolution of PQS in its first minutes of life by three snapshots as~\cite{SteinerPLB00,Shao35,Menezes Influnece, Menezes PNS, Shao11}
\begin{eqnarray}
(\text I)~&S/n_B=1&, Y_l=0.4,\\
(\text {II})~&S/n_B=2&, Y_{\nu_l}=0,\\
(\text {III})~&S/n_B=0&, Y_{\nu_l}=0.
\end{eqnarray}
\section{Results and discussions}

For our numerical calculations, following the Ref. \cite{Chu2014}, the set of
parameters for the current mass of particles we used is: $m_{u0}=m_{d0}=5.5$MeV, $m_{s0}=80$MeV, $m_e=0.511$MeV, and $m_{\mu}=105.7$MeV. We also choose two typical sets of parameters from \cite{Chu2014}: (1) DI-0 with $D_I=0,~D=123.328 \text{MeV~fm}^{-3z},~\alpha=0.7,~ \beta=0.1\text{fm}^{3},~\text{and}~z=1/3$, and (2) DI-85 with $D_I=85 ~\text{MeV~fm}^{-3\alpha},~D=22.922 \text{MeV~fm}^{-3z},~\alpha=0.7,~ \beta=0.1\text{fm}^{3},~\text{and}~z=1.8$, where the former parameter set corresponding to a typical parameter set in the CDDM model \cite{Pen99} is used here for comparison purposes, while the latter parameter set can be used to describe the recently discovered large-mass pulsar PSR J0348+ 0432 with the mass of 2.01$\pm$0.04$M_{\odot}$~\cite{Ant13} as a QS at zero temperature within the CIDDM model\cite{Chu2014}.
\subsection{The quark matter symmetry energy and symmetry free energy}
In the Ref.\cite{Chu2014}, we have considered quark matter within the CIDDM model at zero temperature and found that including isospin dependence of the equivalent quark mass can significantly influence the quark matter symmetry energy as well as the properties of SQM and quark stars. If the scaling parameter $z=1/3$, as the result shows, the equivalent quark mass should be strongly isospin dependent so as to describe PSR J0348+ 0432 as a QS, indicating that the $u$-$d$ quark matter symmetry energy should be much larger than the nuclear matter symmetry energy. On the other hand, if we change the scaling parameter $z$ and set $z=1.8$, the two-flavor quark matter symmetry energy could be smaller than the nuclear matter symmetry energy, while its strength should be at least about twice than that of a free quark gas or quark matter within the conventional Nambu-Jona-Lasinio (NJL) model at the baryon density $1.5\text{fm}^{-3}$ in order to describe PSR J0348+ 0432 as a QS.

In order to investigate the properties of the asymmetric quark matter, the EOS of SQM, and quark stars at finite temperature, it is interesting firstly to study the properties of the symmetry energy and symmetry free energy. As shown in Eq.~(\ref{binding}) and Eq.~(\ref{symfree}), the value of symmetry energy and symmetry free energy would not be equal once finite temperature case is considered. From the Ref.~\cite{Chu2014}, the prediction of the symmetry energy at zero temperature will be very different when $D_I$ changes, which will affect the properties of the quark matter equation of state and QSs largely and show the importance of isospin dependence in asymmetric quark matter within the CIDDM model. One of the main purpose in this work is to investigate the thermodynamical properties of asymmetric quark matter within the CIDDM model at finite temperatures. Then, for the first step, one can study the behaviors of symmetry energy and symmetry free energy in non-zero temperature case by changing isospin dependence with $D_I$.

\begin{figure}[tbp]
\includegraphics[scale=0.43]{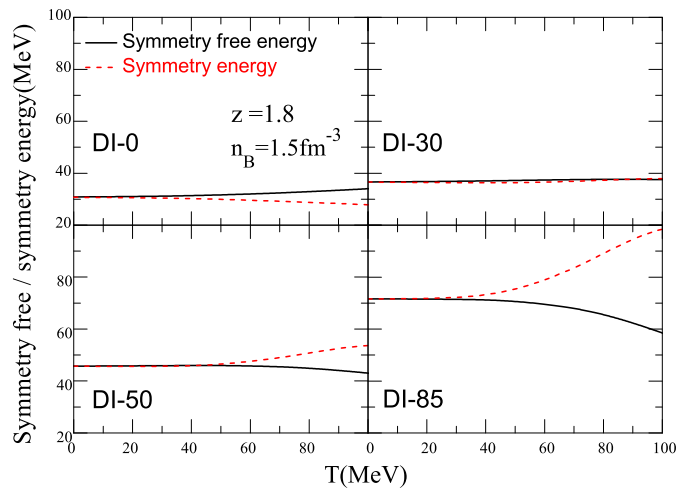}
\caption{(Color online) Two-flavor $u$-$d$ quark matter symmetry energy~{(dashed lines)} and symmetry free energy~{(solid lines)} as functions of temperature in the CIDDM model by using different DI with $z=1.8$, $\alpha=0.7$, and D$=22.922$~MeV~$\text{fm}^{-3z}$, when the baryon density is $1.5~\text{fm}^{-3}$. }
\label{SEFE}
\end{figure}

\begin{figure}[tbp]
\includegraphics[scale=0.42]{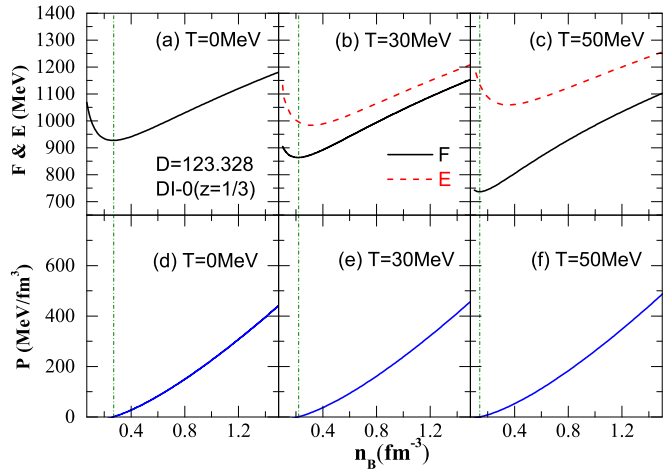}
\caption{(Color online) Energy per baryon {(dashed lines)}, free energy per baryon {(solid lines)} and the corresponding pressure as functions of the baryon density for SQM within the CIDDM model with DI-0 at different temperature. }
\label{EoSDI0}
\end{figure}
In Fig.~\ref{SEFE}, we draw the two-flavor $u$-$d$ quark matter symmetry energy and symmetry free energy as functions of temperature within the CIDDM model. As shown in the work \cite{Chu2014}, the parameter set DI-85 ($z=1.8$) can give the minimum value of $D_I$ (and thus the smallest quark matter symmetry energy) which is necessary to support a QS with a mass of 2.01$M_{\odot}$ in the CIDDM model at zero temperature. Then in Fig.~\ref{SEFE}, we fix the scaling parameter as $z=1.8$ and investigate the properties of the symmetry energy and symmetry free energy with different isospin dependence in the CIDDM model by varying the value of $D_I$. In order to follow the density dependence of the quark matter symmetry energy of a free quark fermi gas or that predicted by the conventional NJL model \cite{Chu2014}, we set $\alpha=0.7$ and $\beta=0.1~$fm$^3$. The baryon density for quark matter in this figure is fixed at 1.5$~\text{fm}^{-3}$, because the central density of the maximum mass configuration for quark stars are usually around 1.5$~\text{fm}^{-3}$ by using the ordinary quark phenomenonlogical models, which will reveal the correlation between the symmetry energy/symmetry free energy and the maximum mass of QSs by using this baryon density.

For the upper left side graph in Fig.~\ref{SEFE}, we set the value of $D_I$ to be zero, which means that there is no isospin dependence in the quark equivalent mass within the CIDDM model. One can see that the value of the symmetry energy and symmetry free energy remain almost unchanged till the temperature reaches 30 MeV. Then the value of symmetry free energy increases with temperature, while the value of symmetry energy decreases, leaving a temperature-dependent splitting between the symmetry energy and symmetry free energy. For DI-30 case, as shown in the upper right side graph in Fig.~\ref{SEFE}, one can find that the values of the symmetry energy and the symmetry free energy stay almost constants. As demonstrated in the lower panels in Fig.~\ref{SEFE}, with $D_I=50~\text{MeV~fm}^{-3\alpha}$ and $D_I=85~\text{MeV~fm}^{-3\alpha}$, the temperature-dependent splitting becomes larger with the increment of temperature, where the value of symmetry energy is higher than that of the symmetry free energy, which demonstrates different properties from that with DI-0. These results indicate that the values of the symmetry energy (symmetry free energy) increases (decreases) or decreases (increases) with the increment of the temperature within the CIDDM model, when $D_I$ is larger (less) than $30~\text{MeV~fm}^{-3\alpha}$ with $z=1.8$, which shows the influence of the isospin dependence in the equivalent quark mass on symmetry energy and symmetry energy at finite temperature.
\subsection{The EOS of SQM}
\begin{figure}[tbp]
\includegraphics[scale=0.42]{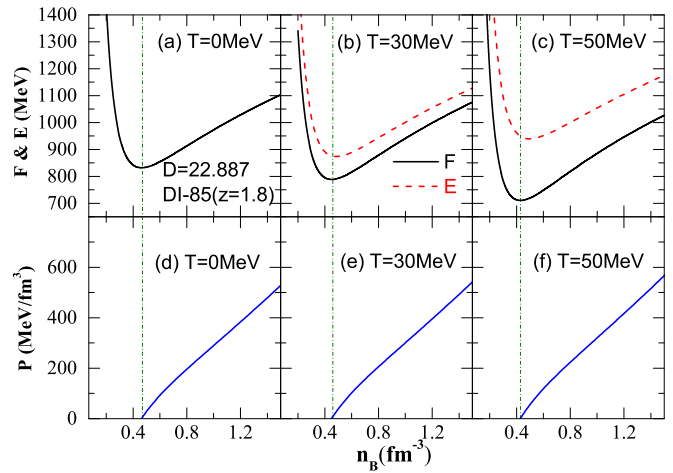}
\caption{(Color online)Energy per baryon {(dashed lines)}, free energy per baryon {(solid lines)} and the corresponding pressure as functions of the baryon density for SQM within the CIDDM model with DI-85 at different temperature. }
\label{EoSDI85}
\end{figure}

In Ref. \cite{Chu2014}, the authors have considered the stability conditions to constrain the value of parameters in the CIDDM model at zero temperature by following the conjecture of Farhi and Jaffe \cite{Far84}. The so-called absolute stability of SQM requires that the minimum energy per baryon of SQM should be less than the minimum energy per baryon of observed nuclei, i.e., M$(^{56}Fe)c^2/56=930~$MeV, and at the same time the minimum energy per baryon of the beta-equilibrium two-flavor $u$-$d$ quark matter should be larger than 930 MeV to be consistent with standard nuclear physics.

In Fig.~$\ref{EoSDI0}$ and Fig.~$\ref{EoSDI85}$, we show the free energy per baryon and energy per baryon for SQM as functions of baryon density within the CIDDM model at different temperatures by using DI-0 and DI-85, where these two sets of parameters both satisfy the demands of the absolute stability of SQM at zero temperature, which put very strong constraints on the choice of the parameters in quark matter models. We can obtain the following information from Fig.~$\ref{EoSDI0}$ and Fig.~$\ref{EoSDI85}$:
1) At zero temperature case for the left side of the two figures, the energy per baryon and free energy per baryon minimum both correspond exactly to the zero pressure point, which is consistent with the HVH theorem (as shown in Eq.~(\ref{Free}) and (\ref{Pressure})) and the thermodynamic self-consistency; 2) As temperature increases, due to the relation $\mathcal{F}_i=\mathcal{E}_i-TS_i$, one can see that from the figures, in $T=30$ MeV and $T=50$ MeV cases, the zero-pressure point is exactly located at the minimum value on every free energy per baryon line, which is also consistent with the HVH theorem and the thermodynamic self-consistency, while the minimum of the energy per baryon and the zero pressure point are generally not the same at finite temperature; 3) Furthermore, it is seen that the density at the minimum of free energy per baryon decreases with the temperature, i.e., when temperature changes from $0$ to $30~\text{MeV}$ and then to $50~\text{MeV}$, it varies from $0.27~\text{fm}^{-3}$ to $0.22~\text{fm}^{-3}$ and then to $0.14~\text{fm}^{-3}$ for Fig.~$\ref{EoSDI0}$, while the density changes from $0.47~\text{fm}^{-3}$ to $0.46~\text{fm}^{-3}$ and then to $0.43~\text{fm}^{-3}$ for Fig.~$\ref{EoSDI85}$; 4) In addition, one can also see from Fig.~$\ref{EoSDI0}$ and Fig.~$\ref{EoSDI85}$ that at a fixed density, the free energy per baryon (energy per baryon) decreases (increases) with the increment of temperature, leading to a clear temperature-dependent splitting between the free energy per baryon and energy per baryon in the finite temperature cases.

\subsection{Quark stars at finite temperatures}
In Ref. \cite{Chu2014}, the authors have calculated the maximum mass of QSs at zero temperature within the CIDDM model by using the parameter sets as DI-0 and DI-85.
Including isospin dependence of the equivalent quark mass, the authors can describe PSR-J038+0432 as a quark star within DI-85 at zero temperature, while, due to the absence of isospin dependence in the quark mass for the CDDM model, the recently discovered large mass pulsars with masses around 2$~M_{\odot}$ cannot be quark stars within DI-0.

Before we describe the three snapshots in time evolution of PQS within the CIDDM model, it is interesting to study the properties of quark stars at finite temperatures. Using this treatment, we do not consider the isentropic stages for PQS, and the main purpose for calculating the properties of QSs at different temperature is to seek out the relation between the EOS of SQM and the maximum mass of QS at finite temperatures, which is useful to reveal the temperature influence over the maximum mass of QSs precisely.

\begin{figure}[tbp]
\includegraphics[scale=0.42]{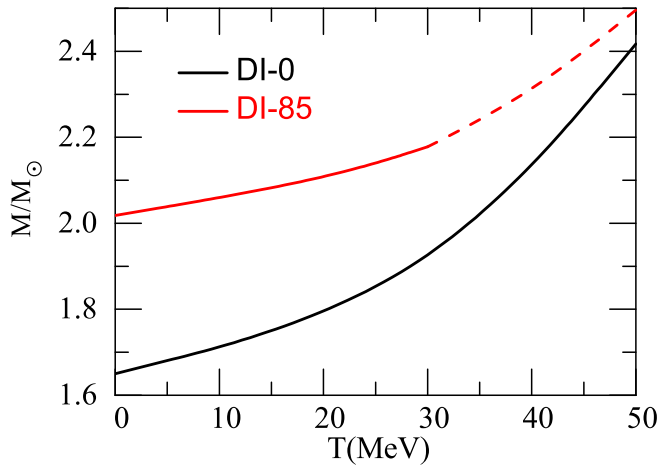}
\caption{(Color online) Maximum mass for static quark stars within the CIDDM model with DI-0 and DI-85 at different temperatures. The dashed line indicates the unphysical region where the EOS with DI-85 violates the causality condition. }
\label{Toy}
\end{figure}
\begin{figure}[tbp]
\includegraphics[scale=0.42]{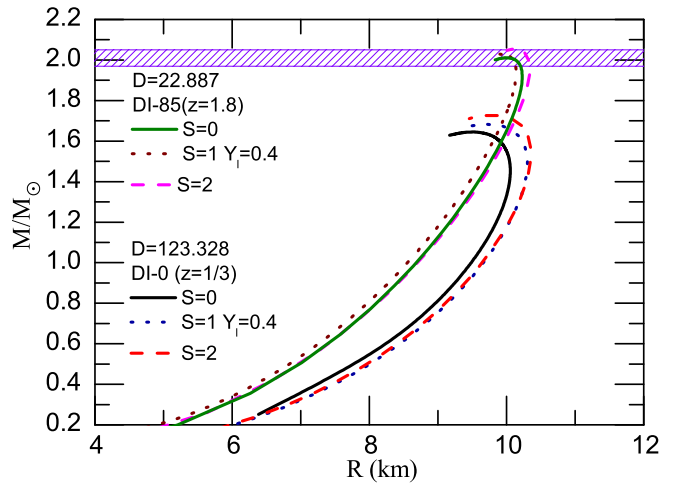}
\caption{(Color online) Mass-radius relations of PQS at three snapshots along a PQS evolution within CIDDM model with DI-0 and DI-85. The shaded band represents the pulsar mass of $2.01\pm 0.04 M_{\odot}$ from PSR J0438+0432\cite{Ant13}.}
\label{RM}
\end{figure}
Shown in Fig.~$\ref{Toy}$ is the maximum mass for static QSs within the CIDDM model with DI-0 and DI-85 at different temperatures. From Fig.~$\ref{Toy}$, one can see that the maximum mass of the static QS is 1.65$M_{\odot}$ at zero temperature with DI-0, which corresponds to the case of the CDDM model, while for the parameter set DI-85, which corresponds to the case of the CIDDM model (using which can describe PSR J0348+0432 as a QS), the maximum mass of the static QS is 2.01$M_{\odot}$ for the zero temperature case. It is obvious to see that the maximum mass of static QSs for both of this two sets of parameters increases with temperature. For the parameter set DI-0, when temperature reaches 50 $MeV$, the maximum mass of the static QS is 2.417 $M_{\odot}$, while for the parameter set DI-85, when temperature reaches 30 $MeV$, the maximum mass for QS is 2.178 $M_{\odot}$. As shown in Fig.~$\ref{Toy}$, the sound velocity with DI-85 will not exceed the velocity of light until the temperature hits 30 $MeV$ {(where the corresponding central baryon density is about 1.11 $\text{fm}^{-3}$)}. Then for the region that the temperature is larger than 30 $MeV$ with DI-85, one should be conscious of that this is the unphysical region for SQM and QS, and then we display this unphysical mass-radius relation with dashed line. The figure also shows that the line of the parameter set DI-0 gives a stronger temperature dependence for the maximum mass of static QSs than that for the parameter set DI-85. From Fig.~$\ref{SEFE}$, the results indicate that the symmetry free energy increases with the increment of temperature within the CIDDM model for the parameter set DI-0, which will provide a stiffer EOS at finite temperature than that from the zero temperature case, while for DI-85, the symmetry free energy decreases with the increment of temperature. Hence the temperature dependence for the maximum mass of the static QS with DI-0 is stronger than that with DI-85.

Since we have calculated the results for the properties of the maximum mass configuration of static QSs within the CIDDM model at finite temperature and found that the maximum mass will increase with temperature, the evolution of PQS should be studied within the CIDDM model.

Shown in Fig.~$\ref{RM}$ is the mass-radius relations of PQS at three snapshots along a PQS evolution within the CIDDM model with DI-0 and DI-85. Firstly, in the heating stage, the PQS expands with the decrement of the lepton fraction and the increment of entropy. When the trapped neutrinos are free, the star begins cooling stage. For the first stage of the evolution of PQS with DI-0, when the entropy per baryon is 1 and the lepton fraction is 0.4, there is a large number of trapped neutrinos, and the maximum mass of PQS is 1.683 $M_{\odot}$. As the trapped neutrinos diffuse, the star continues heating up with the entropy per baryon reaching 2, and then the maximum mass of PQS in the second stage increases to 1.727 $M_{\odot}$, which corresponds to the largest maximum mass case of all the three stages of evolution. Following the heating stages, the star begins cooling and the entropy per baryon reaches zero at the 3rd stage, which is the corresponding zero temperature case, and then the maximum mass of PQS at zero temperature in this snapshot is 1.651 $M_{\odot}$. For DI-85 case, the evolution is similar with that of DI-0: 1) The maximum mass of PQS at the first stage is 2.03 $M_{\odot}$; 2) For the second stage, the maximum mass of PQS increases to 2.053 $M_{\odot}$, which shows the maximum mass case along the evolution; 3) At the 3rd stage, the temperature of the star is cooling down to zero, and then the maximum mass of PQS is 2.01 $M_{\odot}$, which is exactly the center value of the observed mass of $2.01\pm0.04M_{\odot}$ for PSR J0348+0432. These results indicates that the maximum mass of PQS at the heating stages is larger than that of PQS at zero temperature within the CIDDM model with both DI-0 and DI-85, and the maximum mass of PQS in the second stage is always larger than the other two stages, which means that the heating process in the evolution will increase the maximum mass of PQS. Furthermore, one can find that compared with the zero temperature stage, the growth rate of the maximum mass of PQS at the heating stages with DI-0 is larger than that with DI-85, and this phenomenon can also be found in Fig.~$\ref{Toy}$, where the temperature dependence of the maximum mass of the static QS with DI-0 is stronger than that with DI-85.

\section{conclusion and discussion}
In this work, we have extended the CIDDM model, in which the quark confinement is modeled by the density-dependent quark masses, to include temperature dependence of the equivalent quark mass. Within the CIDDM model, we have explored the quark matter symmetry energy, symmetry free energy, the properties of SQM at finite temperatures, and the maximum mass of QSs and PQSs. We have found that including temperature dependence of the equivalent quark mass and considering the temperature effects on the asymmetric quark matter can significantly change the quark matter symmetry energy/symmetry free energy as well as the properties of SQM and stars.

We have demonstrated that within the CIDDM model, the value of the symmetry energy decreases (increases) with the temperature when $D_I$ is less (larger) than $30$, while the symmetry free energy increases (decreases) as the temperature increases when $D_I$ is less (larger) than $30~\text{MeV~fm}^{-3\alpha}$, which can impact differently on the properties of EOS of SQM.

We have further studied the maximum mass of QSs within the CIDDM model at finite temperatures. We have found that both the maximum mass for DI-0 and DI-85 increase with the increment of temperature, and the temperature dependence for the maximum mass of the static QS with DI-0 is stronger than that with DI-85, because the symmetry free energy for DI-0 increases as the temperature increases, which will provide a stiffer EOS for SQM and then increase the maximum mass of QS for DI-0 with a faster pace than the DI-85 case.

Considering three different snapshots along the evolution line of the compact star, we have studied the properties of static PQSs. We have demonstrated that the maximum mass of PQS at the heating stages is always larger than that of PQS at zero temperature within the CIDDM model, which indicates that the heating process in the evolution will increase the maximum mass of PQS.

Therefore, our results have shown that including the temperature dependence of the equivalent quark mass within the CIDDM model can significantly influence the values of symmetry energy, symmetry free energy, properties of the EOS of SQM, and the maximum mass of QSs at finite temperature and PQSs. For the evolution of PQSs, the heating process will increase the maximum mass of the stars.

\vskip 1 cm
{\bf Acknowledgement.---}
This work was supported in part by the Major State Basic Research Development Program (973 Program) in China under
Contract Nos. 2015CB856904 and 2013CB834405, the National Natural Science Foundation of China under Grant
Nos. 11625521, 11275125, 11135011 and 11505100, the Program for Professor of Special Appointment (Eastern Scholar) at Shanghai Institutions of Higher Learning, Key Laboratory for Particle Physics, Astrophysics and Cosmology,Ministry of Education, China, the Science and Technology Commission of Shanghai Municipality (11DZ2260700),
 and the Shandong Provincial Natural Science Foundation, China (ZR2015AQ007).

\end{document}